\documentclass[twocolumn,english,final,showpacs,prl,lengthcheck]{revtex4}
\usepackage[T1]{fontenc}
\usepackage[latin9]{inputenc}
\usepackage{color}
\usepackage{babel}

\usepackage{amsmath}
\usepackage{graphicx}
\usepackage{amssymb}
\usepackage[unicode=true,pdfusetitle,
 bookmarks=true,bookmarksnumbered=false,bookmarksopen=false,
 breaklinks=false,pdfborder={0 0 1},backref=false,colorlinks=true]
 {hyperref}

\makeatletter
\@ifundefined{textcolor}{}
{%
 \definecolor{BLACK}{gray}{0}
 \definecolor{WHITE}{gray}{1}
 \definecolor{RED}{rgb}{1,0,0}
 \definecolor{GREEN}{rgb}{0,1,0}
 \definecolor{BLUE}{rgb}{0,0,1}
 \definecolor{CYAN}{cmyk}{1,0,0,0}
 \definecolor{MAGENTA}{cmyk}{0,1,0,0}
 \definecolor{YELLOW}{cmyk}{0,0,1,0}
 }

\usepackage{textcomp}
\usepackage{babel}

\usepackage{amsfonts}

\providecommand{\U}[1]{\protect\rule{.1in}{.1in}}
\makeatletter
\providecommand{\LyX}{L\kern-.1667em\lower.25em\hbox{Y}\kern-.125emX\@}
\@ifundefined{textcolor}{}
{
\definecolor{BLACK}{gray}{0}
\definecolor{WHITE}{gray}{1}
\definecolor{RED}{rgb}{1,0,0}
\definecolor{GREEN}{rgb}{0,1,0}
\definecolor{BLUE}{rgb}{0,0,1}
\definecolor{CYAN}{cmyk}{1,0,0,0}
\definecolor{MAGENTA}{cmyk}{0,1,0,0}
\definecolor{YELLOW}{cmyk}{0,0,1,0}
}
\makeatother
\makeatother

\makeatother

\usepackage{babel}

\makeatother

\begin{document}

\preprint{This line only printed with preprint option}

\title{Dynamical decoupling noise spectroscopy}

\author{Gonzalo A. \'Alvarez}

\altaffiliation{galvarez@e3.physik.uni-dortmund.de}

\affiliation{Fakult\"at Physik, Technische Universit\"at Dortmund, Dortmund,
Germany.}

\author{Dieter Suter}

\altaffiliation{Dieter.Suter@tu-dortmund.de}

\affiliation{Fakult\"at Physik, Technische Universit\"at Dortmund, Dortmund,
Germany.}


\begin{abstract}
Decoherence is one of the most important obstacles that must be overcome
in quantum information processing. It depends on the qubit-environment
coupling strength, but also on the spectral composition of the noise
generated by the environment. If the spectral density is known, fighting
the effect of decoherence can be made more effective. Applying sequences
of inversion pulses to the qubit system, we generate effective filter
functions that probe the environmental spectral density. Comparing
different pulse sequences, we recover the complete spectral density
function and distinguish different contributions to the overall decoherence.
\end{abstract}
\maketitle
\textit{Introduction.---} Quantum information processing relies on
the robust control of quantum systems. A quantum system is always
influenced by external degrees of freedom, the environment, that disturb
the quantum information by a process called decoherence \cite{Zurek03}.
Many strategies were developed to fight this degradation of information.
These methods are based on correction of errors \cite{3921,6581}
and decoupling the environment \cite{Carr1954,Meiboom1958,Abragam,5916}.
Fighting decoherence successfully requires knowledge of the noise
spectral density to design suitable quantum processes \cite{Zhang2007,Bhaktavatsala2011,Bylander2011,Almog2011}.

One simple decoupling strategy is called dynamical decoupling (DD)
\cite{5916,Yang2010}. It is based on the application of a sequence
of control pulses to the system to effectively isolate it from the
environment. Different DD sequences were developed \cite{5916,khodjasteh_fault-tolerant_2005,Uhrig2008,Yang2010}
and tested experimentally \cite{Biercuk2009a,Du2009,Alvarez2010,Ryan2010,Ajoy2011,Souza2011}.
Different sources modify the performance of DD sequences. Once pulse
errors are small \cite{Alvarez2010,Ryan2010,Souza2011}, the spectral
density of the system-environment (SE) interaction becomes the main
factor \cite{Cywinski2008,Uhrig2008,Gordon2008,Biercuk2009a,Uys2009,Clausen2010,Ajoy2011,Zhang2007}.
Consequently a DD sequence has to be judiciously designed according
to the particular noise spectral density to be decoupled \cite{Cywinski2008,Uhrig2008,Gordon2008,Biercuk2009a,Uys2009,Clausen2010}.
The indefinite number of possibilities for designing DD sequences
leads to the imposibility of a \textquotedblleft{}brute force\textquotedblright{}
exploration and therefore the development of noise spectroscopy methodogies
is required \cite{Meriles2010,Bylander2011,Almog2011,Young2011}.

In this paper, we present a method to determine the spectral density
of the SE interaction. The method is based on previous results that
the decay rate of a qubit during DD is given by the overlap of the
bath spectral density function and a filter function generated by
the DD sequence \cite{Cywinski2008,Uhrig2008,Gordon2008,Biercuk2009a,Uys2009,Clausen2010,Ajoy2011}.
The filter function is given by the Fourier transform of the SE interaction
modified by the control pulses: each $\pi$-pulse changes the sign
of the SE coupling. When many DD cycles are applied to the system,
the filter functions become a sum of $\delta-$functions \cite{Ajoy2011}
and the decoherence time is given by a discrete sum of spectral densities.
A judicious choice of the DD sequence thus allows one to probe the
environmental spectral density at selected frequencies. Combining
several measurements, it is possible to obtain a detailed picture
of the noise spectral distribution. In the following, we describe
an exact and simple method for obtaining general spectral density
functions that extends recent approximate solutions that can be used
only for specific cases \cite{Meriles2010,Bylander2011}.

\textit{A qubit as the noise probe.---} We consider a single qubit
$\hat{S}$ as the probe. It is coupled to the bath to be studied with
a purely dephasing interaction. In a resonantly rotating frame of
reference \cite{Abragam}, the free evolution Hamiltonian is $\widehat{\mathcal{H}}_{f}=\widehat{\mathcal{H}}_{SE}+\widehat{\mathcal{H}}_{E}$,
where $\widehat{\mathcal{H}}_{E}$ is the environment Hamiltonian
and $\widehat{\mathcal{H}}_{SE}=b_{SE}\hat{S}_{z}\hat{E}$ is a general
pure dephasing interaction between system and environment. $\hat{E}$
is some operator of the environment and $b_{SE}$ the SE coupling
strength. This type of interaction is encountered in a wide range
of solid-state spin systems, as for example nuclear spin systems in
NMR \cite{Carr1954,Meiboom1958,Alvarez2010,Ajoy2011}, electron spins
in diamonds \cite{Ryan2010}, electron spins in quantum dots \cite{Hanson2007},
donors in silicon \cite{Kane1998}, etc.

We consider the application of a sequence of short, strong pulses
that invert the probe qubit \cite{Carr1954,Meiboom1958,5916,Yang2010}.
We assume $N$ instantaneous pulses at times $t_{i}$, with delays
$\tau_{i}=t_{i}-t_{i-1}$ between the pulses for $i=2,..,N+1$ and
$\tau_{1}=t_{1}-t_{0}$, where $t_{0}=0$ and $t_{N+1}=\tau_{c}$. 

While such a sequence can refocus a static system-environment interaction
completely, any time-dependence reduces its efficiency. We calculate
the remaining decay rate for the case where the environment can be
well described by a stochastic noise. This results are also valid
for a quantum second order approximation of the time-dependent SE
interaction \cite{Abragam}. We now eliminate the environment-Hamiltonian
$\mathcal{\widehat{H}}_{E}$ by using an interaction representation
with respect to the evolution of the isolated environment. The system-environment
Hamiltonian then becomes $\mathcal{\widehat{H}}_{SE}^{(E)}\left(t\right)=b_{SE}\hat{S}_{z}e^{-i\mathcal{\widehat{H}}_{E}t}\hat{E}e^{i\mathcal{\widehat{H}}_{E}t}$.
Since $\mathcal{\widehat{H}}_{E}$ does not commute with $\mathcal{\widehat{H}}_{SE}$,
the effective system-environment interaction $\mathcal{\widehat{H}}_{SE}^{\left(E\right)}$
is time-dependent and the system experiences a fluctuating coupling
with the environment. Tracing over the bath variables replaces $b_{SE}e^{-i\mathcal{\widehat{H}}_{E}t}\hat{E}e^{i\mathcal{\widehat{H}}_{E}t}$
by the stochastic function $b_{SE}E(t)$. For simplicity we assume
that this random field has a Gaussian distribution with zero average,
$\left\langle E(t)\right\rangle =0$. The auto-correlation function
is $\left\langle E(t)E(t+\tau)\right\rangle =g\left(\tau\right)$
and the spectral density $S(\omega)$ of the system-bath interaction
is the Fourier transform of $b_{SE}^{2}g\left(\tau\right)$.

\begin{figure}
\includegraphics[bb=24bp 135bp 690bp 555bp,clip,width=1\columnwidth]{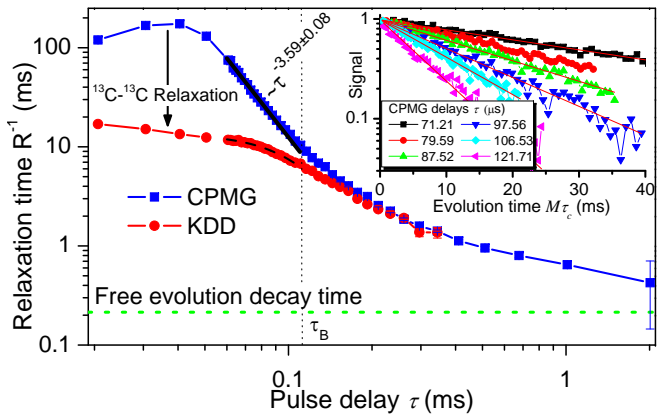}

\caption{(Color online) Experimental relaxation times of the probe spin under
the application of CPMG (squares) and KDD (circles) sequences. The
black solid line represents a power law fitting to the CPMG data and
the green dotted line the asymptotic free evolution decay rate. The
black dashed line is a fitting to the KDD data with an expression
$(R_{13\mathrm{C}}+C^{\prime}\tau^{\alpha})^{-1}$. Inset: Experimental
signal decays of the probe spin as a function of the evolution time
under CPMG dynamical decoupling. Different curves correspond to different
pulse delays. The straight lines represent exponential fits.}
\label{Flo:rates_and_decays}%
\end{figure}

The free evolution operator for a given realization of the random
noise is $\exp\left\{ -i\phi(t)\hat{S}_{z}\right\} $, where $\phi(t)=b_{SE}\int_{0}^{t}dt_{1}E(t_{1})$
is the phase accumulated by the probe spin during the evolution. Considering
now the effect of the pulses, they generate reversals of $\widehat{\mathcal{H}}_{SE}(t)$.
If the pulses are applied during the interval $\tau_{c}$ as described
above, the accumulated phase $\phi(M\tau_{c})$ after $M$ cycles
becomes $\phi(M\tau_{c})=b_{SE}\int_{0}^{M\tau_{c}}dt^{\prime}f_{N}(t^{\prime},M\tau_{c})E(t^{\prime})$,
where the modulating function $f_{N}(\tau^{\prime},M\tau_{c})$ switches
between $\pm1$ at the position of every pulse \cite{Cywinski2008}.
If the initial state of the probe spin is $\hat{\rho}_{0}=\hat{S}_{x,y}$,
its normalized magnetization under the effects of DD taking the average
over the random fluctuations is $\left\langle s_{x,y}(t)\right\rangle =e^{-\frac{1}{2}\left\langle \phi^{2}(t)\right\rangle }$,
and its decay can be quantified by the exponential's argument  \begin{equation}
\tfrac{1}{2}\left\langle \phi^{2}(t)\right\rangle =R(t)\, t=\sqrt{\tfrac{\pi}{2}}\int_{-\infty}^{\infty}d\omega S(\omega)\left|F_{N}(\omega,M\tau_{c})\right|^{2},\label{eq:decaychi}\end{equation}
where $F_{N}(\omega,M\tau_{c})$ is the Fourier transform of $f_{N}(t^{\prime},M\tau_{c})$
\cite{Cywinski2008}. The decay function $R(t)\, t$ is thus equal
to the product of the spectral density $S(\omega)$ of the system-environment
coupling and the filter transfer function $F_{N}(\omega,M\tau_{c})$.
We have recently shown that considering the infinite extension of
the modulating function, $f_{N}(t^{\prime},\infty)$, by a convolution
this provides a $F_{N}(\omega,M\tau_{c})$ that is a sum of sinc functions
centered at the harmonic frequencies $k\omega_{0}=2\pi k/\tau_{c}$
of the Fourier series of $f_{N}(t^{\prime},\infty)$ \cite{Ajoy2011}.
 Hence for $t=M\tau_{c}\gg\tau_{B}$, the noise correlation time,
the filter function $\left|F_{N}(\omega,\tau_{M})\right|$ becomes
an almost discrete spectrum given by the Fourier transform of $f_{N}(t^{\prime},\infty)$,
i.e. $F(\omega,t)$ is represented by a series of $\delta$-functions
centered at $k\omega_{0}$ neglecting the contributions from the
secondary maxima. Thus, in the limit of many cycles, the decay is
exponential and $R(t)$ becomes a constant \begin{equation}
R(t)=R=\sum_{k=1}^{\infty}A_{k}^{2}S\left(k\omega_{0}\right),\label{eq:decMult}\end{equation}
with $A_{k}^{2}=\frac{\sqrt{2\pi}}{\tau_{c}^{2}}\left|F_{N}\left(k\omega_{0},\tau_{c}\right)\right|^{2}$,
where for a CPMG sequence \cite{Carr1954,Meiboom1958} with $\tau_{2}=2\tau_{1}=2\tau_{3}=\tau$,
$A_{k}\propto1/k$ for odd $k$ and 0 otherwise. This is the basis
for the DD noise spectroscopy methodology presented in this letter.
Examples of the probe spin signal decay are shown in the inset of
Fig. \ref{Flo:rates_and_decays}. 

\textit{Noise spectroscopy.--- }Assuming for the moment that the sum
in Eq. (\ref{eq:decMult}) collapses to the $k=1$ term, we can clearly
trace out the bath spectral density by varying the delay between the
pulses as in Ref. \cite{Meriles2010,Bylander2011}. However, for
real DD sequences, we always have an infinite series, where all harmonics
contribute to the decay rate with the weight $A_{k}$. Determining
the spectral density function therefore requires the inversion of
Eq. (\ref{eq:decMult}) and thus the consideration of only the $k=1$
term is a rough approximation. The main difficulty here is that a
single measurement depends on an infinite number of unknown spectral
density values. We solve this problem by a two-step procedure: in
the first step, we combine $m$ measurements with different pulse
delays, which we choose such that they probe the spectral density
function at a discrete set of harmonic frequencies with different
sensitivity amplitudes $A_{k}$. In this step, we neglect contributions
from the tail of $S(\omega>m\omega_{min})$. This yields a square
matrix that we can invert to obtain the values of $S(j\omega_{min}),\, j=1..m$.
From the resulting spectral density function, we estimate a functional
form for the tail of the distribution and correct the data for the
contributions from the tail. Inverting the matrix again, with the
corrected values, gives the final spectral density distribution.

A natural choice for the probing sequence is the CPMG or equidistant
sequence, which has harmonics at frequencies $\omega_{0}=\pi/\tau$.
To simplify inverting equation (\ref{eq:decMult}), we choose the
pulse delays in the different measurements such that all relevant
frequencies, including all harmonics, are multiples of a minimal frequency
$\omega_{min}$. We therefore start with a maximum delay $\tau_{max}=\pi/\omega_{min}$,
which determines the frequency resolution with which we probe the
spectral density function. If the maximum frequency at which we want
to probe the spectral density function directly is $m\omega_{min}$,
then we need to apply sequences with delays $\tau_{n}=\tau_{max}/n=\tau_{min}m/n$.
If we neglect the contribution from frequencies $>m\omega_{min}$,
the relaxation rates $R_{n}$ for the different experiments are given
by a system of $m$ linear equations

\begin{equation}
R_{n}=\sum_{k=1}^{[m/n]}A_{k}^{2}S(nk\omega_{min})=\sum_{j=1}^{m}U_{nj}S_{j},\label{eq:R_cutoff}\end{equation}
where $[m/n]$ denotes the integer part of $m/n$ and $j=nk$. The
elements $A_{k}^{2}$ form an upper triangular matrix $U_{nj}=\sum_{k=1}^{[m/n]}A_{k}^{2}\delta_{j,nk},$
and $S_{j}=S(j\omega_{min})$ represent the unknown spectral density
values, which can formally be calculated as $S_{j}=\sum_{n=1}^{m}\left(U^{-1}\right)_{jn}R_{n}$

We now correct for the omitted contributions from the high-frequency
tail of the infinite sum by approximating it with a suitable functional
form, which depends on the system being studied. Typical examples
include a power law decay, lorentzian or gaussian decay, or a sudden
cut off like in an ohmic bath. In the system that we used as an example
(see below), the experimental data can be approximated very well by
a power law dependence, as shown in Fig. \ref{Flo:rates_and_decays}
(squares).%

If the tail satisfies a power law $S_{j}=\frac{C}{j^{\alpha}}$ for
$j>n_{p}$, then\begin{equation}
R_{n>n_{p}}=\sum_{k=1}^{\infty}\frac{A_{k}^{2}C}{\left(nk\right)^{\alpha}}=\frac{C}{n^{\alpha}}\sum_{k=1}^{\infty}\frac{A_{k}^{2}}{k^{\alpha}}=\frac{C\Lambda_{\alpha}}{n^{\alpha}}.\label{eq:powerlaw}\end{equation}
This relation is represented by the black solid line in Fig. \ref{Flo:rates_and_decays}.
We can now modify Eq. (\ref{eq:R_cutoff}) by adding the neglected
terms and then the relaxation rates satisfy

\begin{equation}
R_{n}=\sum_{j=1}^{m}U_{nj}S_{j}+\left(\frac{\Lambda_{\alpha}C}{n^{\alpha}}-\sum_{j=1}^{m}U_{nj}\frac{C}{j^{\alpha}}\right),\label{eq:exact}\end{equation}
where $\left(\frac{\Lambda_{\alpha}C}{n^{\alpha}}-\sum_{j=1}^{m}U_{nj}\frac{C}{j^{\alpha}}\right)=\frac{C}{n^{\alpha}}\sum_{k>m-n+1}\frac{A_{k}^{2}}{k^{\alpha}}$
represents the effective spectral density summing the contribution
from all harmonics $k>m-n+1$. The spectral density is now determined
from $S_{j}=\sum_{n=1}^{m}\left(U^{-1}\right)_{jn}\left(R_{n}-\frac{\Lambda_{\alpha}C}{n^{\alpha}}\right)+\frac{C}{j^{\alpha}}$.
Eq. (\ref{eq:powerlaw}) shows that for a power law dependence, the
relaxation rate and the spectral density are proportional and thus
for a qualitative description of $S(\omega)$ considering only the
$k=1$ term is enough validating the results of Ref. \cite{Meriles2010,Bylander2011}. 

\textit{Experimental determination of $S(\omega)$.--- }For an experimental
demonstration of this method, we chose $^{13}$C nuclear spins ($S=1/2$)
as probe qubits. We used polycrystalline adamantane where the carbon
nuclear spins are coupled to an environment of $^{1}$H nuclear spins
($I=1/2$) that act as a spin-bath. The natural abundance of the $^{13}$C
nuclei is about 1\%, and to a good approximation each $^{13}$C nuclear
spin is surrounded by about 150 $^{1}$H nuclear spins. The interaction
with the environment is thus dominated by the $^{13}$C-$^{1}$H magnetic
dipole coupling \cite{Abragam}. To determine the bath spectral density
we applied the equidistant sequences CPMG and KDD \cite{Souza2011}
to the probe spin for different delays between pulses $\tau_{n}=\tau_{max}/n$,
with $n=1..40$ and $\tau_{max}=2$ms. Delays are measured between
the center of the pulses. For CPMG, we chose an initial state longitudinal
to the rf field of the refocusing pulses because then pulse error
effects can be neglected \cite{Meiboom1958,Alvarez2010}. The inset
of Fig. \ref{Flo:rates_and_decays} shows examples of the $^{13}$C
signal decays. The lines in the inset show the fitted exponential
decays, which agree very well with the data points in this range.
This demonstrates that we are in the regime where the filter functions
are discrete. KDD was shown to be robust against pulse errors, independent
of the initial condition (please see Ref. \cite{Souza2011} for details).
For ideal pulses, both sequences have the same filter function. As
shown in Fig. \ref{Flo:rates_and_decays} (squares) the observed relaxation
times for this system depend on the pulse spacing like $\propto\tau^{-3.59}$
for the CPMG sequence over the range $\tau=$$[50\,\mu s,110\,\mu s]$.
We only used the data points for $\tau>50\,\mu\mathrm{s}$ to determine
the parameters $C$ and $\alpha$, since Fig. \ref{Flo:rates_and_decays}
indicates that other processes contribute to the relaxation at shorter
delays. From the fitting process, we found $\alpha=3.59\pm0.08$ and
$\Lambda_{\alpha}\approx1.002$. Here, the contribution of the infinite
series of $2\cdot10^{-3}$ is  almost negligible. Figure \ref{Flo:rates_and_decays}
also shows that the dependence of the decoherence rates changes at
$\tau\gtrsim100\,\mu\mathrm{s}$. This agrees with the value that
we determined earlier for the correlation time of the bath $\tau_{B}$
\cite{Alvarez2010}.

We observe in the KDD case that the relaxation time saturates for
$\tau$ shorter than $50\,\mu$s and in general is lower than the
CPMG cases (Fig. \ref{Flo:rates_and_decays}, circles). This difference
can be attributed to the effect of $^{13}$C-$^{13}$C couplings.
Because in the CPMG sequence all pulses generate the same rotation,
the overall effect of the pulse cycle is to first order equivalent
to a constant effective field, which stabilizes the observable magnetization
against the effect of $^{13}$C-$^{13}$C couplings \cite{Santyr1988}.
In the KDD case, the state is not longitudinal to the pulses and no
spin-lock effect is observed. The saturation of the relaxation time
for the CPMG case for $\tau<50\,\mu$s can be attributed to the finite
rf field strength or, equivalently, to the finite duration of the
pulses. Pulse errors may also contribute in this regime.

\begin{figure}
\includegraphics[bb=35bp 150bp 670bp 540bp,clip,width=1\columnwidth]{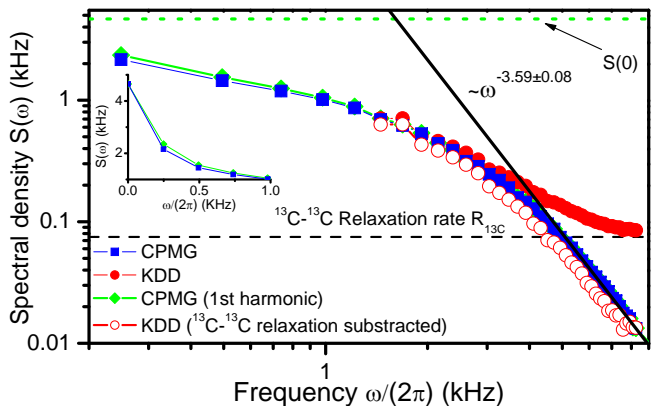}

\caption{(Color online) Experimentally determined noise spectral density. The
inset shows in a linear scale the low frequency regime. The green
dotted line represents the free evolution decay rate of the probe
spin, i.e. $S(0)$.}
\label{Flo:Spectral density}%
\end{figure}

In the KDD case, the $^{13}$C-$^{13}$C interaction eclipses the
spectral density of the proton bath. To verify this, we assumed a
$^{13}$C-$^{13}$C relaxation rate independent of the pulse delays
and fitted the expression $(R_{13\mathrm{C}}+C^{\prime}\tau^{\alpha})^{-1}$
over the range $[50\,\mu s,110\,\mu s]$ where the CPMG data follow
a power law (dashed line in Fig. \ref{Flo:rates_and_decays}). We
obtained $R_{13\mathrm{C}}=\left(75\pm1\right)\,\mathrm{s^{-1}}$
for the $^{13}$C-$^{13}$C relaxation rate and $\alpha=3.7\pm0.3$,
which perfectly matches with the CPMG result. If we subtract the $R_{13\mathrm{C}}$
contribution, we obtain the spectral density represented by the empty
circles in Fig. \ref{Flo:Spectral density}, which are almost identical
to the result obtained with the CPMG sequence (solid squares).

We demonstrate with Eq. (\ref{eq:powerlaw}) that the qualitative
behaviour of the power law tail can be well obtained by the first
harmonic approximation ($k=1$) that is proportional to the exact
solution derived from (\ref{eq:exact}). For lower frequencies, the
corrections from our exact method can be relevant (squares vs. rombuses
in Fig. \ref{Flo:Spectral density}). However, because $S(\omega)$
decays rapidly, the difference is small in this case. 

\begin{figure}
\includegraphics[bb=40bp 150bp 670bp 580bp,clip,width=1\columnwidth]{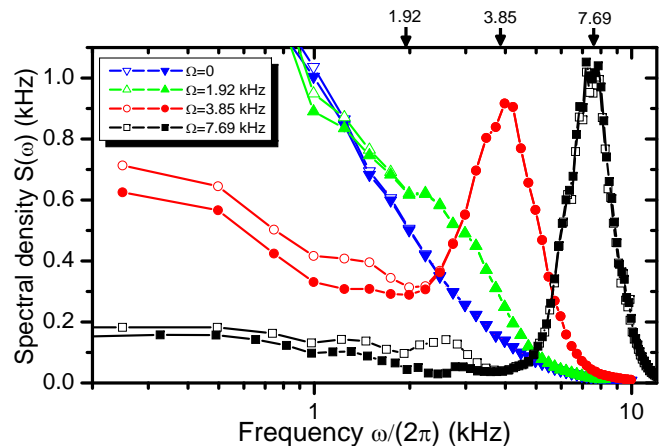}

\caption{(Color online) Experimentally determined noise spectral density for
a modulated system-environment interaction. The empty symbols were
obtained when only the first harmonic is considered, the full symbols
show the results from the exact method.}
\label{Flo:Sw-CW}%
\end{figure}

This is not always true, as we now show with a specific example: We
modulate the system-environment interaction by applying a resonant
radio-frequency field to the proton spins, which periodically inverts
them. Using the same measurement procedure, we obtained the data shown
in Fig. \ref{Flo:Sw-CW}. 

The approximate solution $S(n\omega_{min})=R_{n}/A_{1}^{2}$, where
only the first harmonic ($k=1$) is considered, shows a main peak
at the modulation frequency $\Omega$ and some satellite peaks at
at lower frequencies that are integer fractions of $\Omega$. These
satellite peaks are artefacts of the data analysis that neglects contributions
from higher harmonics of the filter function of Eq. (\ref{eq:decMult}).
Using our method, we obtain an improved solution where the satellite
peaks are eliminated. The spectral density distribution for this case
is qualitatively different from that of the unmodulated case. In particular,
the value at zero frequency is reduced, but a maximum has appeared
at the modulation frequency. This has important consequences for implementing
dynamical decoupling: A good decoupling performance when $\Omega=7.69$
kHz is expected for $\tau\sim0.12$ ms, where the first harmonic is
at the spectral density minimum near 4 kHz. Increasing the decoupling
rate then would drastically reduce the decoupling performance, in
stark contrast to the usual expectation that it should increase with
the decoupling rate.

\textit{Conclusions.---} We have developed a method to determine the
noise spectral density generated by a bath. It is based on modulating
the system-environment interaction by applying sequences of inversion
pulses to the system. If the sequence consists of many repetitions
of a basic cycle, the resulting decays are exponentials and the decay
rates are given by the spectral density at discrete frequencies. This
allows one to build a linear system of equations that can be inverted
to obtain the unknown spectral density function. We applied the method
to obtain the spectral density of the $^{13}$C- $^{1}$H interaction
in adamantane. Applying this method to other systems will help fighting
decoherence, e.g. by optimizing DD sequences by reducing the overlap
of their filter functions with the noise spectral density \cite{Cywinski2008,Uhrig2008,Gordon2008,Biercuk2009a,Uys2009,Clausen2010}.
Our method also complements standard NMR techniques that use CPMG
sequences to distinguish between difference sources of inhomogenities
\cite{Garroway1977} or measuring diffusion rates \cite{Stejskal1965a,Stejskal1965,Packer1973}
as well as protein dynamics rates \cite{Mittermaier2006} in liquid
state NMR. Those methods determine correlation times but assume specific
spectral density functions, while our technique is suitable for the
determination of unknown spectral densities.
\begin{acknowledgments}
Acknowledgments.--- This work is supported by the DFG through Su 192/24-1.
GAA thanks financial support from the Alexander von Humboldt Foundation
in the initial stage of this project. We thank Alexandre M. Souza
for helpful discussions.
\end{acknowledgments}
\bibliographystyle{apsrev} \bibliographystyle{apsrev}
\bibliography{DDnoisespect,bibliography,CDD,CDD2}

\begin{thebibliography}{34}
\expandafter\ifx\csname natexlab\endcsname\relax\def\natexlab#1{#1}\fi
\expandafter\ifx\csname bibnamefont\endcsname\relax
  \def\bibnamefont#1{#1}\fi
\expandafter\ifx\csname bibfnamefont\endcsname\relax
  \def\bibfnamefont#1{#1}\fi
\expandafter\ifx\csname citenamefont\endcsname\relax
  \def\citenamefont#1{#1}\fi
\expandafter\ifx\csname url\endcsname\relax
  \def\url#1{\texttt{#1}}\fi
\expandafter\ifx\csname urlprefix\endcsname\relax\def\urlprefix{URL }\fi
\providecommand{\bibinfo}[2]{#2}
\providecommand{\eprint}[2][]{\url{#2}}

\bibitem[{\citenamefont{Zurek}(2003)}]{Zurek03}
\bibinfo{author}{\bibfnamefont{W.~H.} \bibnamefont{Zurek}},
  \bibinfo{journal}{Rev. Mod. Phys.} \textbf{\bibinfo{volume}{75}},
  \bibinfo{pages}{715} (\bibinfo{year}{2003}).

\bibitem[{\citenamefont{Preskill}(1998)}]{3921}
\bibinfo{author}{\bibfnamefont{J.}~\bibnamefont{Preskill}},
  \bibinfo{journal}{Proc. R. Soc. Lond. A} \textbf{\bibinfo{volume}{454}},
  \bibinfo{pages}{385} (\bibinfo{year}{1998}).

\bibitem[{\citenamefont{Knill}(2005)}]{6581}
\bibinfo{author}{\bibfnamefont{E.}~\bibnamefont{Knill}},
  \bibinfo{journal}{Nature} \textbf{\bibinfo{volume}{434}}, \bibinfo{pages}{39}
  (\bibinfo{year}{2005}).

\bibitem[{\citenamefont{Carr and Purcell}(1954)}]{Carr1954}
\bibinfo{author}{\bibfnamefont{H.~Y.} \bibnamefont{Carr}} \bibnamefont{and}
  \bibinfo{author}{\bibfnamefont{E.~M.} \bibnamefont{Purcell}},
  \bibinfo{journal}{Phys. Rev.} \textbf{\bibinfo{volume}{94}},
  \bibinfo{pages}{630} (\bibinfo{year}{1954}).

\bibitem[{\citenamefont{Meiboom and Gill}(1958)}]{Meiboom1958}
\bibinfo{author}{\bibfnamefont{S.}~\bibnamefont{Meiboom}} \bibnamefont{and}
  \bibinfo{author}{\bibfnamefont{D.}~\bibnamefont{Gill}},
  \bibinfo{journal}{Rev. Sci. Instrum.} \textbf{\bibinfo{volume}{29}},
  \bibinfo{pages}{688} (\bibinfo{year}{1958}).

\bibitem[{\citenamefont{Abragam}(1961)}]{Abragam}
\bibinfo{author}{\bibfnamefont{A.}~\bibnamefont{Abragam}},
  \emph{\bibinfo{title}{Principles of Nuclear Magnetism}}
  (\bibinfo{publisher}{Oxford University Press, London}, \bibinfo{year}{1961}).

\bibitem[{\citenamefont{Viola et~al.}(1999)\citenamefont{Viola, Knill, and
  Lloyd}}]{5916}
\bibinfo{author}{\bibfnamefont{L.}~\bibnamefont{Viola}},
  \bibinfo{author}{\bibfnamefont{E.}~\bibnamefont{Knill}}, \bibnamefont{and}
  \bibinfo{author}{\bibfnamefont{S.}~\bibnamefont{Lloyd}},
  \bibinfo{journal}{Phys. Rev. Lett.} \textbf{\bibinfo{volume}{82}},
  \bibinfo{pages}{2417} (\bibinfo{year}{1999}).

\bibitem[{\citenamefont{Zhang et~al.}(2007)\citenamefont{Zhang, Peng,
  Rajendran, and Suter}}]{Zhang2007}
\bibinfo{author}{\bibfnamefont{J.}~\bibnamefont{Zhang}},
  \bibinfo{author}{\bibfnamefont{X.}~\bibnamefont{Peng}},
  \bibinfo{author}{\bibfnamefont{N.}~\bibnamefont{Rajendran}},
  \bibnamefont{and} \bibinfo{author}{\bibfnamefont{D.}~\bibnamefont{Suter}},
  \bibinfo{journal}{Phys. Rev. A} \textbf{\bibinfo{volume}{75}},
  \bibinfo{pages}{042314} (\bibinfo{year}{2007}).

\bibitem[{\citenamefont{Bhaktavatsala~Rao and
  Kurizki}(2011)}]{Bhaktavatsala2011}
\bibinfo{author}{\bibfnamefont{D.~D.} \bibnamefont{Bhaktavatsala~Rao}}
  \bibnamefont{and} \bibinfo{author}{\bibfnamefont{G.}~\bibnamefont{Kurizki}},
  \bibinfo{journal}{Phys. Rev. A} \textbf{\bibinfo{volume}{83}},
  \bibinfo{pages}{032105} (\bibinfo{year}{2011}).

\bibitem[{\citenamefont{Bylander et~al.}(2011)\citenamefont{Bylander,
  Gustavsson, Yan, Yoshihara, Harrabi, Fitch, Cory, Nakamura, Tsai, and
  Oliver}}]{Bylander2011}
\bibinfo{author}{\bibfnamefont{J.}~\bibnamefont{Bylander}},
  \bibinfo{author}{\bibfnamefont{S.}~\bibnamefont{Gustavsson}},
  \bibinfo{author}{\bibfnamefont{F.}~\bibnamefont{Yan}},
  \bibinfo{author}{\bibfnamefont{F.}~\bibnamefont{Yoshihara}},
  \bibinfo{author}{\bibfnamefont{K.}~\bibnamefont{Harrabi}},
  \bibinfo{author}{\bibfnamefont{G.}~\bibnamefont{Fitch}},
  \bibinfo{author}{\bibfnamefont{D.~G.} \bibnamefont{Cory}},
  \bibinfo{author}{\bibfnamefont{Y.}~\bibnamefont{Nakamura}},
  \bibinfo{author}{\bibfnamefont{J.}~\bibnamefont{Tsai}}, \bibnamefont{and}
  \bibinfo{author}{\bibfnamefont{W.~D.} \bibnamefont{Oliver}},
  \bibinfo{journal}{Nat Phys} \textbf{\bibinfo{volume}{7}},
  \bibinfo{pages}{565} (\bibinfo{year}{2011}).

\bibitem[{\citenamefont{Almog et~al.}(2011)\citenamefont{Almog, Sagi, Gordon,
  Bensky, Kurizki, and Davidson}}]{Almog2011}
\bibinfo{author}{\bibfnamefont{I.}~\bibnamefont{Almog}},
  \bibinfo{author}{\bibfnamefont{Y.}~\bibnamefont{Sagi}},
  \bibinfo{author}{\bibfnamefont{G.}~\bibnamefont{Gordon}},
  \bibinfo{author}{\bibfnamefont{G.}~\bibnamefont{Bensky}},
  \bibinfo{author}{\bibfnamefont{G.}~\bibnamefont{Kurizki}}, \bibnamefont{and}
  \bibinfo{author}{\bibfnamefont{N.}~\bibnamefont{Davidson}},
  \bibinfo{journal}{J. Phys. B: At., Mol. Opt. Phys.}
  \textbf{\bibinfo{volume}{44}}, \bibinfo{pages}{154006}
  (\bibinfo{year}{2011}).

\bibitem[{\citenamefont{Yang et~al.}(2010)\citenamefont{Yang, Wang, and
  Liu}}]{Yang2010}
\bibinfo{author}{\bibfnamefont{W.}~\bibnamefont{Yang}},
  \bibinfo{author}{\bibfnamefont{Z.}~\bibnamefont{Wang}}, \bibnamefont{and}
  \bibinfo{author}{\bibfnamefont{R.}~\bibnamefont{Liu}},
  \bibinfo{journal}{Front. Phys.} \textbf{\bibinfo{volume}{6}},
  \bibinfo{pages}{2} (\bibinfo{year}{2010}).

\bibitem[{\citenamefont{Khodjasteh and
  Lidar}(2005)}]{khodjasteh_fault-tolerant_2005}
\bibinfo{author}{\bibfnamefont{K.}~\bibnamefont{Khodjasteh}} \bibnamefont{and}
  \bibinfo{author}{\bibfnamefont{D.~A.} \bibnamefont{Lidar}},
  \bibinfo{journal}{Phys. Rev. Lett.} \textbf{\bibinfo{volume}{95}},
  \bibinfo{pages}{180501} (\bibinfo{year}{2005}).

\bibitem[{\citenamefont{Uhrig}(2008)}]{Uhrig2008}
\bibinfo{author}{\bibfnamefont{G.~S.} \bibnamefont{Uhrig}},
  \bibinfo{journal}{New J. Phys.} \textbf{\bibinfo{volume}{10}},
  \bibinfo{pages}{083024} (\bibinfo{year}{2008}).

\bibitem[{\citenamefont{{M. J. Biercuk, \emph{et al.}}}(2009)}]{Biercuk2009a}
\bibinfo{author}{\bibnamefont{{M. J. Biercuk, \emph{et al.}}}},
  \bibinfo{journal}{Nature} \textbf{\bibinfo{volume}{458}},
  \bibinfo{pages}{996} (\bibinfo{year}{2009}).

\bibitem[{\citenamefont{{J. Du, \emph{et al.}}}(2009)}]{Du2009}
\bibinfo{author}{\bibnamefont{{J. Du, \emph{et al.}}}},
  \bibinfo{journal}{Nature} \textbf{\bibinfo{volume}{461}},
  \bibinfo{pages}{1265} (\bibinfo{year}{2009}).

\bibitem[{\citenamefont{\'{A}lvarez et~al.}(2010)\citenamefont{\'{A}lvarez,
  Ajoy, Peng, and Suter}}]{Alvarez2010}
\bibinfo{author}{\bibfnamefont{G.~A.} \bibnamefont{\'{A}lvarez}},
  \bibinfo{author}{\bibfnamefont{A.}~\bibnamefont{Ajoy}},
  \bibinfo{author}{\bibfnamefont{X.}~\bibnamefont{Peng}}, \bibnamefont{and}
  \bibinfo{author}{\bibfnamefont{D.}~\bibnamefont{Suter}},
  \bibinfo{journal}{Phys. Rev. A} \textbf{\bibinfo{volume}{82}},
  \bibinfo{pages}{042306} (\bibinfo{year}{2010}).

\bibitem[{\citenamefont{Ryan et~al.}(2010)\citenamefont{Ryan, Hodges, and
  Cory}}]{Ryan2010}
\bibinfo{author}{\bibfnamefont{C.~A.} \bibnamefont{Ryan}},
  \bibinfo{author}{\bibfnamefont{J.~S.} \bibnamefont{Hodges}},
  \bibnamefont{and} \bibinfo{author}{\bibfnamefont{D.~G.} \bibnamefont{Cory}},
  \bibinfo{journal}{Phys. Rev. Lett.} \textbf{\bibinfo{volume}{105}},
  \bibinfo{pages}{200402} (\bibinfo{year}{2010}).

\bibitem[{\citenamefont{Ajoy et~al.}(2011)\citenamefont{Ajoy, \'{A}lvarez, and
  Suter}}]{Ajoy2011}
\bibinfo{author}{\bibfnamefont{A.}~\bibnamefont{Ajoy}},
  \bibinfo{author}{\bibfnamefont{G.~A.} \bibnamefont{\'{A}lvarez}},
  \bibnamefont{and} \bibinfo{author}{\bibfnamefont{D.}~\bibnamefont{Suter}},
  \bibinfo{journal}{Phys. Rev. A} \textbf{\bibinfo{volume}{83}},
  \bibinfo{pages}{032303} (\bibinfo{year}{2011}).

\bibitem[{\citenamefont{Souza et~al.}(2011)\citenamefont{Souza, \'{A}lvarez,
  and Suter}}]{Souza2011}
\bibinfo{author}{\bibfnamefont{A.~M.} \bibnamefont{Souza}},
  \bibinfo{author}{\bibfnamefont{G.~A.} \bibnamefont{\'{A}lvarez}},
  \bibnamefont{and} \bibinfo{author}{\bibfnamefont{D.}~\bibnamefont{Suter}},
  \bibinfo{journal}{Phys. Rev. Lett.}  \textbf{\bibinfo{volume}{106}},
  \bibinfo{pages}{240501} (\bibinfo{year}{2011}).

\bibitem[{\citenamefont{Cywinski et~al.}(2008)\citenamefont{Cywinski, Lutchyn,
  Nave, and DasSarma}}]{Cywinski2008}
\bibinfo{author}{\bibfnamefont{L.}~\bibnamefont{Cywinski}},
  \bibinfo{author}{\bibfnamefont{R.~M.} \bibnamefont{Lutchyn}},
  \bibinfo{author}{\bibfnamefont{C.~P.} \bibnamefont{Nave}}, \bibnamefont{and}
  \bibinfo{author}{\bibfnamefont{S.}~\bibnamefont{DasSarma}},
  \bibinfo{journal}{Phys. Rev. B} \textbf{\bibinfo{volume}{77}},
  \bibinfo{pages}{174509} (\bibinfo{year}{2008}).

\bibitem[{\citenamefont{Gordon et~al.}(2008)\citenamefont{Gordon, Kurizki, and
  Lidar}}]{Gordon2008}
\bibinfo{author}{\bibfnamefont{G.}~\bibnamefont{Gordon}},
  \bibinfo{author}{\bibfnamefont{G.}~\bibnamefont{Kurizki}}, \bibnamefont{and}
  \bibinfo{author}{\bibfnamefont{D.~A.} \bibnamefont{Lidar}},
  \bibinfo{journal}{Phys. Rev. Lett.} \textbf{\bibinfo{volume}{101}},
  \bibinfo{pages}{010403} (\bibinfo{year}{2008}).

\bibitem[{\citenamefont{Uys et~al.}(2009)\citenamefont{Uys, Biercuk, and
  Bollinger}}]{Uys2009}
\bibinfo{author}{\bibfnamefont{H.}~\bibnamefont{Uys}},
  \bibinfo{author}{\bibfnamefont{M.~J.} \bibnamefont{Biercuk}},
  \bibnamefont{and} \bibinfo{author}{\bibfnamefont{J.~J.}
  \bibnamefont{Bollinger}}, \bibinfo{journal}{Phys. Rev. Lett.}
  \textbf{\bibinfo{volume}{103}}, \bibinfo{pages}{040501}
  (\bibinfo{year}{2009}).

\bibitem[{\citenamefont{Clausen et~al.}(2010)\citenamefont{Clausen, Bensky, and
  Kurizki}}]{Clausen2010}
\bibinfo{author}{\bibfnamefont{J.}~\bibnamefont{Clausen}},
  \bibinfo{author}{\bibfnamefont{G.}~\bibnamefont{Bensky}}, \bibnamefont{and}
  \bibinfo{author}{\bibfnamefont{G.}~\bibnamefont{Kurizki}},
  \bibinfo{journal}{Phys. Rev. Lett.} \textbf{\bibinfo{volume}{104}},
  \bibinfo{pages}{040401} (\bibinfo{year}{2010}).

\bibitem[{\citenamefont{Meriles et~al.}(2010)\citenamefont{Meriles, Jiang,
  Goldstein, Hodges, Maze, Lukin, and Cappellaro}}]{Meriles2010}
\bibinfo{author}{\bibfnamefont{C.~A.} \bibnamefont{Meriles}},
  \bibinfo{author}{\bibfnamefont{L.}~\bibnamefont{Jiang}},
  \bibinfo{author}{\bibfnamefont{G.}~\bibnamefont{Goldstein}},
  \bibinfo{author}{\bibfnamefont{J.~S.} \bibnamefont{Hodges}},
  \bibinfo{author}{\bibfnamefont{J.}~\bibnamefont{Maze}},
  \bibinfo{author}{\bibfnamefont{M.~D.} \bibnamefont{Lukin}}, \bibnamefont{and}
  \bibinfo{author}{\bibfnamefont{P.}~\bibnamefont{Cappellaro}},
  \bibinfo{journal}{J. Chem. Phys.} \textbf{\bibinfo{volume}{133}},
  \bibinfo{pages}{124105} (\bibinfo{year}{2010}).

\bibitem[{\citenamefont{Young and Whaley}(2011)}]{Young2011}
\bibinfo{author}{\bibfnamefont{K.~C.} \bibnamefont{Young}} \bibnamefont{and}
  \bibinfo{author}{\bibfnamefont{K.~B.} \bibnamefont{Whaley}},
  \bibinfo{journal}{arXiv:1102.5115}  (\bibinfo{year}{2011}).

\bibitem[{\citenamefont{{R. Hanson, \emph{et al.}}}(2007)}]{Hanson2007}
\bibinfo{author}{\bibnamefont{{R. Hanson, \emph{et al.}}}},
  \bibinfo{journal}{Rev. Mod. Phys.} \textbf{\bibinfo{volume}{79}},
  \bibinfo{pages}{1217} (\bibinfo{year}{2007}).

\bibitem[{\citenamefont{Kane}(1998)}]{Kane1998}
\bibinfo{author}{\bibfnamefont{B.~E.} \bibnamefont{Kane}},
  \bibinfo{journal}{Nature} \textbf{\bibinfo{volume}{393}},
  \bibinfo{pages}{133} (\bibinfo{year}{1998}).

\bibitem[{\citenamefont{Santyr et~al.}(1988)\citenamefont{Santyr, Henkelman,
  and Bronskill}}]{Santyr1988}
\bibinfo{author}{\bibfnamefont{G.~E.} \bibnamefont{Santyr}},
  \bibinfo{author}{\bibfnamefont{R.~M.} \bibnamefont{Henkelman}},
  \bibnamefont{and} \bibinfo{author}{\bibfnamefont{M.~J.}
  \bibnamefont{Bronskill}}, \bibinfo{journal}{J. Magn. Reson.}
  \textbf{\bibinfo{volume}{79}}, \bibinfo{pages}{28} (\bibinfo{year}{1988}).

\bibitem[{\citenamefont{Garroway}(1977)}]{Garroway1977}
\bibinfo{author}{\bibfnamefont{A.~N.} \bibnamefont{Garroway}},
  \bibinfo{journal}{J. Magn. Reson.} \textbf{\bibinfo{volume}{28}},
  \bibinfo{pages}{365} (\bibinfo{year}{1977}).

\bibitem[{\citenamefont{Stejskal}(1965)}]{Stejskal1965a}
\bibinfo{author}{\bibfnamefont{E.~O.} \bibnamefont{Stejskal}},
  \bibinfo{journal}{J. Chem. Phys.} \textbf{\bibinfo{volume}{43}},
  \bibinfo{pages}{3597} (\bibinfo{year}{1965}).

\bibitem[{\citenamefont{Stejskal and Tanner}(1965)}]{Stejskal1965}
\bibinfo{author}{\bibfnamefont{E.~O.} \bibnamefont{Stejskal}} \bibnamefont{and}
  \bibinfo{author}{\bibfnamefont{J.~E.} \bibnamefont{Tanner}},
  \bibinfo{journal}{J. Chem. Phys.} \textbf{\bibinfo{volume}{42}},
  \bibinfo{pages}{288} (\bibinfo{year}{1965}).

\bibitem[{\citenamefont{Packer}(1973)}]{Packer1973}
\bibinfo{author}{\bibfnamefont{K.~J.} \bibnamefont{Packer}},
  \bibinfo{journal}{J. Magn. Reson.} \textbf{\bibinfo{volume}{9}},
  \bibinfo{pages}{438} (\bibinfo{year}{1973}).

\bibitem[{\citenamefont{Mittermaier and Kay}(2006)}]{Mittermaier2006}
\bibinfo{author}{\bibfnamefont{A.}~\bibnamefont{Mittermaier}} \bibnamefont{and}
  \bibinfo{author}{\bibfnamefont{L.~E.} \bibnamefont{Kay}},
  \bibinfo{journal}{Science} \textbf{\bibinfo{volume}{312}},
  \bibinfo{pages}{224} (\bibinfo{year}{2006}).

\end{thebibliography}

\end{document}